\documentstyle[11pt]{article}
\newcommand{\blankline}{\vskip .3cm}
\newcommand{\f}{\begin{equation}}
\newcommand{\ff}{\end{equation}}

\newcommand{\ttau}{\rlap{\lower2ex\hbox{$\,\tilde{}$}}\tau{}}
\newcommand{\tG}{\rlap{\lower2ex\hbox{$\,\,\tilde{}$}}{G}}
\newcommand{\tepsilon}{\rlap{\lower2ex\hbox{$\,
\tilde{}$}}\epsilon{}}
\newcommand{\tH}{\rlap{\lower2ex\hbox{$\,\,\tilde{}$}}{H}}
\newcommand{\tN}{\rlap{\lower2ex\hbox{$\,\,\tilde{}$}}{N}}
\begin{document}
\rightline{ gr-qc/9405015 , CGPG-94/4-1}
\centerline{\LARGE  The Chern-Simons invariant
as the natural time}
\blankline
\centerline{\LARGE  variable for classical and quantum cosmology}
\blankline
\rm
\vskip.2cm
\centerline{Lee Smolin${}^*$ and Chopin Soo ${}^\dagger$}
\blankline
 \centerline{\it  Center for Gravitational Physics and Geometry}
\centerline{\it Department of Physics}
 \centerline {\it The Pennsylvania State University}
\centerline{\it University Park, PA 16802, USA}
 \blankline
\centerline{(Revised and extended)}
\blankline
\centerline{\today}
\blankline
\centerline{ABSTRACT}

We propose that the Chern-Simons invariant of the Ashtekar-Sen
connection is the natural
internal time coordinate for  classical and quantum cosmology.
The reasons for this are a number
of interesting properties of this functional, which we describe here.
1)It is a function on the
 gauge
and diffeomorphism invariant configuration space, whose gradient is
orthogonal
 to the two physical degrees of freedom, in the metric defined by the
 Ashtekar
 formulation of general relativity. 2)The imaginary part of the
Chern-Simons
form  reduces in the limit of small cosmological constant,
$\Lambda$, and
solutions close to DeSitter spacetime, to the York extrinsic time
 coordinate.
3)Small matter-field excitations of the Chern-Simons  state
satisfy, by
virtue of the quantum constraints, a functional Schroedinger
 equation in which
 the matter fields evolve on a DeSitter background in the
Chern-Simons time.
We then propose this is the natural vacuum state of the theory
for $\Lambda
\neq 0$. 4)This time coordinate is periodic on the configuration
space of
Euclideanized spacetimes, due to the large gauge
transformations, which means
that physical expectation values for all states in non-perturbative
quantum
gravity will satisfy the $KMS$ condition, and may then be
 interpreted as
thermal states. 5)Forms for the physical hamiltonians and inner
product which support the proposal are suggested, and a new action
 principle
for general relativity, as a geodesic principle on the connection
superspace,
is found.

\vfill
${}^*$ smolin@phys.psu.edu \quad ${}^\dagger$ soo@phys.psu.edu
\eject

\section{Introduction}

To construct a single unified physical theory from general relativity
and
quantum field theory we must be able to extend quantum theory to
the
universe as a whole.   But efforts to accomplish this have
so far failed at least
in part because of the
problem of time\cite{problem-of-time}.  This stems from the
apparent conflict between, on the one hand, the
quantum theory's need to refer to a
preferred clock when defining the notions of evolution and exclusive
outcomes that are essential for the probability interpretation and, on
the other hand,
 the diffeomorphism invariance of general relativity, which forbids
description in terms of a fixed or absolute notion of time, external to
the universe.  One proposal that has been made about this problem is
that we might be able to identify a degree of freedom of the
gravitational
field that could serve as a clock, with respect to which the evolution
of
both
matter and the other gravitational degrees of freedom
might be measured\cite{BSW,york-internal,karel-internal}.
In this letter we would like to propose that there is a particularly
natural
choice for such an ``internal" clock made from the geometry of
spacetime, it
is the the Chern-Simons invariant\cite{cs}
of the Ashtekar-Sen
connection, $A_{ai}$.
Our proposal applies both to physical, Lorentzian
spacetimes and to their Euclidean extensions, which are expected
to be useful for
the quantum theory.

Before stating our proposal, there is an important point about the
specification of the time function that must be stressed.  In some
treatments of the problem of time, it is stated that what is required
is
a time function on the spacetime manifold
itself\cite{problem-of-time}.  This must be a
scalar
function on spacetime, satisfying certain properties.  This may be a
useful
thing for the classical theory, but it suffers from a severe limitation
when
we attempt to approach the problem of time in the quantum theory,
which is
that, in general, the spacetime has no meaning in the quantum
theory.
Instead, the natural arena for the quantum theory is the
configuration space.
As is emphasized in many applications of quantum mechanics, the
configuration
space is the place on which the quantum states are defined, and in
which their
evolution takes place. More than this, spacetime
can be
expected to play no fundamental role in the interpretation of
quantum gravity, just as trajectories of particles play no role
in  the interpretation of
ordinary quantum mechanics.  It may emerge in the classical limit,
but
for the exact, non-perturbative theory, quantum states will exist on
the
configuration space of general relativity, and they will not normally
have
a simple interpretation in terms of spacetime.

For this reason, what is wanted to address the problem of time in the
quantum
theory is a time function on the configuration space.  Now, it might
seem at
first that it is possible to have both, that is to have a local function on
spacetime
that, given any slicing into three dimensional surfaces, can be
integrated to
give a time function on the space of configurations.  However, a little
thought
leads to the conclusion that this is not possible, for the simple reason
that if a
function is going to be integrated over a three manifold to give a
time function
on the configuration space, it must be a density on that three surface.
In that case
it cannot come from a function on spacetime without the specification
of additional
information.  Given an appropriate spatial density, one can make
such an
association, but as any such density must be, in a diffeomorphism
invariant
field theory,
a dynamical degree of freedom, the difference is significant.

Thus, we have to make a choice as to whether the object we are
investigating
is a time function on the configuration space or a time function on a
spacetime.
For the reasons stated, we choose the former. Finally, we may
emphasize that
from the point of view of the quantum theory,
one does not need to have specified a slicing condition to speak of the
configuration
space in terms of functions on an abstract three
manifold.  Instead,
the slicing
condition is to be seen as merely a gauge condition that helps, when
appropriate,  to translate the fundamental
dynamics on the configuration space into statements about
spacetimes\footnote{Of course, there are
cases where it is interesting to employ a gauge fixing
as an auxiliary device to describe a particular spacetime in terms
of a (gauge dependent) trajectory in the configuration space,
rather than in terms of a gauge-equivalent class of trajectories.
In these cases a gauge
condition that chooses the
slicings must be specified. This is completely compatible with what
we do here.
In such cases our proposal becomes a way to label the slices picked
out by
the slicing condition. For some applications this may not be
necessary,
but if one is interested in studying the dynamics of the gravitational
field, it may
be natural to use a label on the slices that measures dynamical
information
about the gravitational field, rather than simply the gauge dependent
information
coded in the slicing condition.}.

Having specified the context, we may now discuss what we claim and
do in
this paper.
The main result of the Ashtekar formalism, and all the work done
using it, is
to reinforce the idea that it is useful to think of the configuration
space of
general relativity as built from an $SO(3)$
connection on an abstract
three manifold.  The main idea of this paper is that it may be useful
for
certain purposes to conceive of time as being measured by
a particular function on this configuration space, which is the Chern-
Simons invariant
of that connection.

The evidence we have for this conjecture comes from both the physical regime of
the theory where the metric has a Minkowskian
signature and the Euclidean regime, which is relevant for the
path integral formulation of the theory and for the discussion
of the thermodynamics of the gravitational field.   In the latter
case, the Ashtekar-Sen connection, $A_{ai}$, is real and we propose to take
\f
\tau_{CS} =
\int_\Sigma  Y_{CS}  (A)
\ff
as a measure of the Euclideanized time.
Here $Y_{CS} (A) = {1\over 2}(A^i \wedge dA^i + {1 \over 3}
\epsilon_{ijk}
A^i \wedge A^j \wedge A^k) $ is the
Chern-Simons form\footnote{We use the notation
in which $a,b,c,...$ are spatial indices and $i,j,k$ are internal $SO(3)$
indices that label the frame fields of space.  We may note
that we are taking here the Ashtekar connection to have
natural dimensions of inverse length, which means it differs
from the convention usually employed by a factor of
Newton's constant, $G$.
We will use units here in which $\hbar =1$, but $G$ is
written explicitly, so that $G$ has dimensions of $(length)^2$,
while the cosmological constant $\Lambda$ has dimensions
of $(length)^{-4}$. The combination $\lambda=G^2\Lambda $, where
$G$ is Newton's constant, is then dimensionless.},
which is integrated over a spatial three manifold, $\Sigma$, which,
as we are studying cosmology, we take to be compact.

As we shall see, the evidence that the Chern-Simons invariant
plays the role of a time coordinate is, in the Euclidean case,
rather direct, and comes from an analysis of both the geometry
of the configuration space and the hamiltonian formulation of
the theory.
In the physical, Minkowskian case, the situation is
more complicated because the
Ashtekar-Sen connection, $A_a^i$, is
complex.  This means that the whole Chern-Simons invariant
is complex, and cannot directly serve as a measure of time,
in the classical theory, in the same way it appears to
in the Euclidean case.   However,
we are able to show that
in the semiclassical limit,
and  in the case of a non-vanishing
cosmological constant,  the imaginary part of the
Chern-Simons invariant
\f
\tau_{ICS}=  Im
\int_\Sigma  Y_{CS}(A)   ,
\ff
does plays the role of a time coordinate
for the theory.

The evidence for our proposal, which is described in the
remainder of this paper, may then be summarized as follows.

1) In the configuration space of left-handed spin
connections appropriate to the Ashtekar formulation
of general relativity\cite{abhay-reviews}, the Chern-Simons
invariant, $\int Y_{CS} $, is a
natural time coordinate in that its gradient is both timelike in
connection superspace (for Euclidean signature spacetimes),
and orthogonal to variations in the two
physical degrees of freedom,
with
respect to the natural metric on the configuration
space\cite{chopin-chang}.
This natural metric arises from the Hamiltonian constraint
of the Ashtekar formulation.
In Sections 2-4 we proceed to investigate and
substantiate this claim, for the case of vanishing cosmological
constant.
We show in Section 2 that the geometry of the connection superspace
can be understood in a simple way in terms of structures
derived from the Chern-Simons functional.  This leads
us, in Section 3,
to the discovery of a simple action principle, which
expresses the fact, previously known, that for vanishing
cosmological constant the classical spacetimes correspond to
a certain class of null geodesics of the metric on the
connection configuration space.  We find that
the Chern-Simons functional is the natural candidate for
the parametrization of trajectories and null geodesics in the
arena of connection superspace.  We then find, in
Section 4, and
again for the case of vanishing cosmological constant,
that the Hamiltonian which evolves the gravitational field
in the Chern-Simons form has a simple form as the square
root of a positive functional.  This allows us to derive a
simple Schroedinger equation for
the evolution of the quantum state on configuration space in
the Chern-Simons time\footnote{We find also an alternative
form of the time parameter, as a kind of averaged Chern-Simons
time (eq. 35), which gives an extremely simple
form for a functional Schroedinger equation.}.
It also suggests that
``stationary states" with respect to the Chern-Simons time may exist
in non-perturbative quantum gravity.

2) When the cosmological constant is non-vanishing the Hamiltonian
constraint is cubic in the momentum conjugate to the Ashtekar-Sen
connection. As a result, the use of the Chern-Simons functional as an
{\it exact} choice for ``time" may no longer be valid.
In Sections 5 and 6, we discuss how the Chern-Simons functional
can still emerge as a natural choice for ``time", both in the presence
and absence of matter, when the cosmological constant $\lambda$ is
non-vanishing but small.

In Section 5 we consider the semiclassical regime of the
theory.
By considering states of the form
$\Psi[A,\phi] = \psi_{CS}[A]\chi[A, \phi]$, where $\phi$
is a matter field and
$\psi_{CS}$, the Chern-Simons
state (eqn. 37, below), is
an exact quantum state in the absence of matter\cite{kodama},
we show that in the limit of
small $\lambda$ and solutions that are close to DeSitter spacetime,
the
quantum constraints reduce to a
Schroedinger equation in which $\chi$ evolves in a time given by
$K$, the trace of the extrinsic curvature $K_{ai}$.   Further, in this
same limit it is easy to show that
\f
Y_{CS}= \imath {\lambda \over 3G} \sqrt{det(g)}K
 +O(\sqrt{\lambda} )
\ff
so that the times as measured by $\sqrt{g}K$ and by the
imaginary part of the Chern-Simons form
coincide.  Futhermore, the real part of the $Y_{CS}$ is of higher
order, so that it becomes purely imaginary in this limit.  It is as a
result of this that we are able to assert that
in the semiclassical limit of the Minkowskian signature theory,
the imaginary part of the Chern-Simons invariant is playing the
role of the time coordinate.
We may recall that the trace of the
extrinsic curvature was proposed some time ago by York to be the
internal time of general relativity\cite{york-internal}  and was
studied
by Kuchar \cite{karel-internal} and others in the context
of several models as well as in the semiclassical limit.
We believe that $\tau_{ICS}$ offers a natural way to preserve what
is useful about the York time coordinate, while extending it in
a way that gives a simple description of the non-perturbative
dynamics of the theory.

3) This choice of a physical time then leads to a particular form
for the physical inner
product, described in Section 7, which may be computable in
terms of a power series in $K_{ai}$, when use is made of Witten's
discovery\cite{witten-cs} of the connection
between Chern-Simons theory and the Jones polynomial of knot
theory.

4) Because it takes the imaginary part, $\tau_{ICS}$ is invariant
under large gauge transformations, which modify only the real
part of the Chern-Simons functional.
However, if we continue to the Euclidean theory then, due to the
large gauge
transformations, $\tau_{CS}=\int_{\Sigma}Y_{CS}$
 becomes a periodic variable on the now real
connection configuration space.
In Section 8, we argue that, as a result,
expectation values defined in terms of path
integral over the connection-configuration space
will satisfy the KMS condition\cite{kms}, so that
all states of quantum cosmology must be thermal.
This extends the results found earlier for the semiclassical theory
around
classical spacetime solutions\cite{hartle-hawking,GHaw},
and may offer a useful perspective on the link between ``time" and
thermodynamics in non-perturbative quantum gravity.

The paper closes with a concluding section that summarizes the
different results found here.

\section{The geometry of the connection-configuration space}

We begin by describing the
geometry of the kinematical connection configuration space which is
the
space of connections, $\cal A$, modulo the space of all $SO(3)$-
valued small
gauge transformations, $\cal G$,
\f
{\cal C}_{kin} = { {\cal A}/{\cal G} }
\ff
and the related
diffeomorphism invariant configuration space
\f
{\cal C}_{diff} = {\cal C}_{kin} /Diff(\Sigma) .
\ff

We will take the viewpoint here that
classical general relativity as a dynamical theory is
best understood in terms of trajectories in ${\cal C}_{diff}$.
It is then important to use whatever information can be
gained about the geometry of this space.  Here we describe
what we know about the natural metric structure of this space
from previous work\cite{chopin-chang} and extend that analysis.  We may
note
that many things we would like to know are not yet well studied,
among these
are many interesting global questions that we will not be able to
address.
The results we describe here for connection superspace may be
compared with
the geometry of the configuration space of three metrics studied first
by DeWitt\cite{bryce}.

For simplicity we also restrict our discussion
of the geometry
of ${\cal C}_{diff}$ to the Euclidean signature
case.   We also assume throughout this paper that the
spatial manifold $\Sigma$ is compact.

We may note that we do not know good coordinates for the
configuration space ${\cal C}_{diff}$.  For this reason we begin
using the coordinates on the space of connections $\cal A$ on
$\Sigma$,
and then investigate explicitly how to go down to the moduli spaces
${\cal C}_{kin}$ and ${\cal C}_{diff}$.

Our starting point is to notice that $\tau_{CS}$ does describe a
functional on
${\cal C}_{diff}$ and to investigate the consequences of choosing it
for a time function.  We then seek to decompose the geometry of
the infinite dimensional configuration spaces locally by making a
splitting of the geometry according to this time function, analogous to
the usual $3+1$ splitting we make in classical relativity.

Thus, the gradient of the time function is given by\footnote{As usual,
densities are often denoted by tildes.  We may note that as we are
on the space of connections, the indices are reversed, so that an
abstract cotangent space index on ${\cal A}$
is composed of a spatial coordinate, an internal
$SO(3)$ index and a spatial one form index.}
\f
\tilde{\tau}^{ai} (x) \equiv {\delta \tau_{CS} \over \delta
A_{ai}(x)}
= {1 \over 2}{\tilde \epsilon}^{abc} F_{bc}^i
\ff
where $F_{bc}^i$ is the curvature of the Ashtekar-Sen connection.
Much of the simplicity of what follows is a consequence of this
fact that
the left-handed spacetime curvature
may be interpreted to be precisely the gradient of a natural time
function on the configuration space $\cal A$.  There is
a natural contravariant metric on ${\cal C}_{kin}$, which is
\f
\tG_{ai,bj}= {1 \over \tilde{b}} \tepsilon_{abc}
\epsilon_{ijk} \tilde{\tau}^{ck} .
\ff
Here, $\tilde{b} =  (det\tilde{\tau}^{ai})^{1/2}$
is necessary so that $\tG_{ai,bj}$ has
the appropriate density weight on $\Sigma$ to be the contravariant
metric
on ${\cal C}_{kin}$.  We note that when $\tilde{b}=0$ the metric on
${\cal C}_{kin}$ becomes degenerate.  For the purposes of this paper
we
assume we are away from such points, which are in any case non-
generic.

We may note, further, that (7) is special, in that it is
diagonal in the points of
the spatial manifold $\Sigma$.   A general contravariant metric
on the configuration space would be of the form of
$G_{ai,bj}(x,y)$.  In these terms we might write (7) as
$G_{ai,bj}(x,y)=\tilde{b}^{-1} \tepsilon_{abc}
\epsilon_{ijk} \tilde{\tau}^{ck} \delta^3 (y,x)$.

Using the contravariant metric, we may contract with the index of
the
gradiant of the time function, to find the unit time vector field,
\f
\ttau_{ai} \equiv {1 \over 6\tilde{b}} \tG_{ai , bj}{\tilde \tau}^{bj} =
{1 \over  6\tilde{b}^2 }
\tepsilon_{abc}\epsilon_{ijk}
{\tilde\tau}^{bj}{\tilde \tau}^{ck}
\ff
It follows directly that at each point of
$\Sigma, \ \ $    $\ttau_{ai}{\tilde \tau}^{ai} = 1 $.
Note that we choose
this definition for the unit time vector field because our aim is
to describe structures that are local in space as well as in
the configuration space.

Using these functions we may then find the metric on the slices of
${\cal C}_{kin}$ of constant $\tau_{CS}$, for example
the projection operator into the slices of constant $\tau_{CS} $ is
\f
H_{ai}^{\ \ \ bj} \equiv \ttau_{ai}{\tilde \tau^{bj}}-
\delta_a^b\delta_i^j
\ff
while lowering with $\tG_{ai, bj}$ gives
\f
\tH_{ai, bj} \equiv 6\tilde{b} \ttau_{ai} \ttau_{bj} - \tG_{ai , bj}
\ff
The full configuration space metric may then be written
\begin{eqnarray}
ds^2_{superspace} &=& \int_\Sigma \tilde{G}^{ai, bj}
\delta A_{ai} \delta A_{bj}
\nonumber \\
&=& \int_{\Sigma} \left ( {1\over  {6\tilde b}}({\tilde \tau}^{ai}\delta
A_{ai})^2 -
  \tilde{H}^{ai , bj} \delta A_{ai} \delta A_{bj} \right ),
\end{eqnarray}
where the covariant connection superspace metric is
\f
\tilde{G}^{ai, bj} (x) =
{1 \over \tilde{b}}
\left ( {1 \over 2} \tilde{\tau}^{ai} \tilde{\tau}^{bj}
- \tilde{\tau}^{bi} \tilde{\tau}^{aj}    \right ) .
\ff
Note that $\int \tilde{\tau}^{ai}\delta A_{ai} = \delta \int Y_{CS}$, so
that
\f
\tilde{H}^{ai, bj} = {1\over 6{\tilde b}}
{\tilde\tau}^{ai}{\tilde\tau}^{bj} -\tilde{G}^{ai,bj}
\ff
 is the metric on constant
$\tau_{CS}$ surfaces.

Now, from previous work\cite{chopin-chang}, we know that
for Euclidean
spacetimes,
${\tilde H}^{ai,bj}$ has signature $(5, 3)\times \infty^3$,
where the three negative
directions at each point
correspond to changes of the connections under
spatial diffeomorphisms,
and
the space spanned by the five positive directions at each point
include three gauge
degrees which correspond to $SO(3)$ gauge transformations.
Thus, after gauge fixing, $\tilde{G}^{ai,bj}$ may be pulled back to a
signature $(1,2) \times \infty^3$ metric on the
diffeomorphism invariant configuration space, ${\cal C}_{diff}$.
We see that the gradiant of $\tau_{CS}$ spans the $\infty^3$
``timelike" directions.  The remaining $2\times \infty^3$
``spacelike" directions, orthogonal to $\tilde{\tau}^{ai}$,
must then be
considered to be the variations in the
two physical degrees of freedom of the gravitational field.

We may note that when the connection is complex, the same
decomposition of the degrees of freedom may be done.  However
it no longer is meaningful to talk of the signature of the
configuration space metric.  Whether it is meaningful when
the Lorentzian reality conditions are imposed, which
involve relations between configuration and momenta
variables, is presently an open problem.

\section{Einstein's theory from a geodesic principle
on the connection configuration space}

We may note that the usual Ashtekar form of the
Hamiltonian constraint is
\f
{\cal H}= \tilde{b}\tG_{ai, bj} \tilde{E}^{ai}\tilde{E}^{bj}
+ {\lambda \over 3G}det( {\tilde{E}}^{ai}) = 0
\ff
Note that when  $\lambda=0$ what is relevant is only
the  ``conformal class"
of metrics on ${\cal C}_{kin}$ that differ from ${\tilde G}^{ai,bj}$
by the multiplication by a free, nonvanishing
function, $\Omega$, of $\Sigma$.  Thus, for $\lambda =0$,
physical spacetimes are defined by the common geodesics of the
conformal class of metrics ${\Omega}{\tilde G}^{ai,bj}$ on ${\cal
C}_{kin}$.
These may be called the ``locally-null" geodesics of ${\tilde
G}^{ai,bj}$.
Another way to say this is that a simple action
principle for general relativity, written  in terms of
connections is \footnote{This
is related to
the Capovilla-Dell-Jacobson action\cite{CDJ}.  It is interesting
to note that, in contrast to the Barlein, Sharp and Wheeler form
of the action \cite{BSW}, the shifts do not appear explicitly.
Instead the diffeomorphism constraints appear by
requireing the closure of the algebra generated by
the Hamiltonian
constraint.},
\f
S_{GR}[A, dA, N] \equiv { 1\over 16\pi{G}}\int dt
{\int_\Sigma { 1\over N}\tilde{G} ^{ai,bj}\dot{A}_{ai}\dot{A}_{bj}}
\ff
{}From this action, the momentum conjugate to $A_{ai}$ is given by
\f
{\tilde E}^{ai}/(8 \pi{G}) =  \tilde{G}^{ai, bj}\dot{A}_{bj}/({N}8\pi{G})
\ff
Rewritten in the canonical form, the action is
\f
S_{GR}[A, {\tilde E}, N] \equiv {1 \over 8\pi{G}}\int dt
{\int_\Sigma {\tilde E}^{ai}\dot{A}_{ai} - ({\tN}/2)
 {\tG}_{ai, bj}
{\tilde E}^{ai}{\tilde E}{bj}}
\ff
Variation with respect to $\tN$ results in the locally null geodesic
constraint
\f
{\tilde G}^{ai, bj}\dot{A}_{ai}\dot{A}_{bj} = 0
\ff
or equivalently (as we have said earlier, we assume ${\tilde b} \neq
0$)
\f
{\rlap{\lower2ex\hbox{$\,\,\tilde{}$}}{G}_{ai,bj}}
{\tilde E}^{ai}{\tilde E}^{bj} = 0
\ff
All of the dynamics and constraints of general relativity are
derived from this simple functional (expression (15)) of the
left-handed spin connection. By taking the Poisson bracket of the
``scalar" (superhamiltonian) constraint (19) with itself, the ``vector"
(supermomentum) constraint emerges as a secondary constraint. The
Poisson
bracket of the vector constraints then yields Gauss' Law as a further
constraint. Thus the locally null geodesic constraint reproduces
the rest of the Ashtekar constraints of 3d-diffeomorphism and
$SO(3)$
gauge invariance through secondary constraints and the requirement
of
closure. With the action (15), under t-reparametrization
$t \mapsto T$,
$1/N$ scales by $dt/dT$. Since $1/{N}$ is a Lagrange multiplier,
the
physics is t-reparametrization invariant. A natural and explicitly
gauge
and diffeomorphism invariant choice for $T$ as a parametrization of
null
geodesics in connection-superspace is precisely
$\int_\Sigma Y_{CS}$ \footnote{
Equivalently, we may write the action principle in a way that is
explicitly
invariant under reparametrizations of the t coordinate on the
trajectories
in the configuration space.  Just as we do for ordinary geodesics
on finite dimensional manifolds, we may
write the action principle as
\f
S_{GR}^\prime [A, dA, \Omega ] \equiv { 1\over 16\pi{G}}\int dt
\sqrt{\int_\Sigma  \Omega \tilde{G}^{ai,bj}\dot{A}_{ai}\dot{A}_{bj} } .
\ff
Of course, this variational principle is subject to the same
difficulty of the standard variation principle for null geodesics,
which is that the canonical momenta diverge when the
equations of motion (18) are satisfied.  However, it is still
the case that the constraint (19) is satisfied, as may be
seen by rescaling the momenta by the square root
of the constraint (19).  Once this is done, one finds that
variation of this action by $\Omega$
then leads to the locally null
geodesic constraint (18), and the Hamiltonian constraint (19),
and hence to the full Hamiltonian formulation of general
relativity.}.

Finally, before we close this section, it is interesting to note that this
construction generalizes to give a connection between a large class
of topological topological
field theories in three dimensions and a class of gravitational
theories in $3+1$ dimensions.  Because
any three manifold
supplies us with a natural $\tepsilon_{abc}$, any functional,
$f(A)$ on
${\cal C}_{diff}$ determines an inverse metric on  that space,
and hence by the geodesic principle on ${\cal C}_{diff}$
a dynamics for the
gravitational field.
Our construction may then be generalized by replacing $\tau_{CS}$
by any functional $f(A)$ on a space of connections on a three
manifold.
The result is that any topological field theory on a three manifold,
specified by an action $S(A)$, then corresponds through eqn. (7)
to a gravitational theory given by the action (15).\footnote{
Note that when the group is larger than $SO(3)$ the
condition that the algebra of constraints closes requires
that the inverse
metric (7) may be written in the Peldan form\cite{peldan}
in which
$\epsilon^{ijk} \rightarrow \tepsilon_{abc}\tilde{B}^{ai}
\tilde{B}^{bj}\tilde{B}^{ck}/det(\tilde{B})$.}
This may be
related
to the theories studied in \cite{peldan}.

\section{A canonical transformation and a hamiltonian}

Given the proposal that the Chern-Simons functional is
to be regarded as labeling surfaces of constant ``time"
in the configuration space, as well as
a natural parametrization of trajectories there,
we would like to ask if it is possible to make a coordinate
transformation
to exhibit explicitly a decomposition of the directions in the
configuration
space into ``timelike'' and physical degrees of freedom.
If this can be done we
would further like to know if we can perform a
canonical
transformation on the phase space to exhibit that splitting.
What we will show here is that, at least in the Euclidean
case, this can be done locally, in the phase space.
One byproduct of this
will be that this canonical transformation will yield,
at least for the case of vanishing cosmological constant, a
Hamiltonian
that generates the evolution of the gravitational
degrees of freedom in $\tau_{CS}$.

We begin by looking for coordinates on the
constant $\tau_{CS}$ sections of ${\cal C}_{kin}$.
To find these, it is natural to split the variations in
the connection $A_{ai}$ into components that are parallel
to, and orthogonal to, the ``timelike" direction $\ttau_{ai}$.
We then define,
\f
\delta  A_{ai} = \ttau_{ai} \delta \tilde{A}^{||} - \delta A^\perp_{ai}
\ff
where,
\f
\delta \tilde{A}^{||} \equiv \tilde{B}^{ai}\delta A_{ai}
\ff
and
\f
\delta A^\perp_{ai} \equiv H_{ai}^{\ \ \ bj}\delta A_{bj}
= - \delta A_{ai} +
\ttau_{ai}{\tilde \tau}^{bj}\delta A_{bj}
\ff
Note that here we use the common shorthand,
$\tilde{B}^{ai}= {1 \over 2} \epsilon^{abc}F_{bc}^i$.
It of course follows that
$\tilde{\tau}^{ai}\delta A_{ai}^\perp =0$.
\footnote{It is interesting to note that $A^\perp$ must
transform as a
connection
under the restricted algebra of
$\tau_{CS}$-time independent gauge transformations, which
are defined by gauge parameters $\Lambda$ that preserve
its orthogonality to $\tau^{ai}$ so that
$
0={\tilde \tau}^{bj}  \delta_{\Lambda} A_{bj}
= {\tilde\tau}^{bj} {\cal D}_b \Lambda_j .
$
It may then be useful to express the theory in terms of loop
variables
associated with this new connection.}.

It would be very useful to be able  to integrate these equations,
to define $\tilde{A}^{||}$ and $A^\perp_{ai}$ as coordinates on
the configuration
space.  We have not investigated the questions of what
global obstructions there may be to the definition of
such coordinates, or the related question of what initial conditions
may be used to define solutions to the differential relations (22-23).
However, even without this, certain things can be said.
First, it follows that,
\f
(\ast 1)\delta \tilde{A}^{||} = (\delta Y_{CS})
-{1\over 2}d(\delta A^{i} \wedge A^i)
\ff
where $(\ast 1)$ is the volume element. So
\f
\int_\Sigma \delta \tilde{A}^{||}(\ast 1)= \int_\Sigma \delta Y_{CS}
\ff
from which it follows that up to an overall constant
\f
\int_\Sigma \tilde{A}^{||}(\ast 1)= \int_\Sigma  Y_{CS}
\ff

Even without information on the global existence of the
coordinates $\tilde{A}^{||}$ and $A^\perp_{ai}$, we can go ahead
and construct a canonical transformation to define momenta
conjugate to them.  Thus, we define a canonical transformation
such that,
\begin{eqnarray}
A_{ai} &\rightarrow &  (\tilde{A}^{||}, A^\perp_{ai} )
\nonumber \\
\tilde{E}^{ai}  & \rightarrow &  (p^{||} , \tilde{p}_\perp^{ai})
\end{eqnarray}
It will be important to note in what follows that it is
$\tilde{A}^{||}$ and $\tilde{p}_\perp^{ai}$ that are the densities.
The
transformation that accomplishes this may
be found from the condition that it must leave the symplectic
structure on the phase space fixed.  In particular, we have
\f
-{{1}\over {8\pi G}}\int_{\Sigma}d^3x{\tilde E}^{ai}\delta A_{ai} =
\int_\Sigma d^3x \left ( p^{||}(x) \tilde{\tau}^{ai}\delta A_{ai}(x) +
\tilde{p}^{ai}_\perp (x) \delta A^\perp_{ai} (x)\right )
\ff
One then finds directly that,
\f
-({1\over 8\pi{G}})\tilde{E}^{ai} (x) =
p^{||}(x) \tilde{\tau}^{ai} + {\tilde p}_\perp^{ck}  H_{ck}^{\ \ \ ai}
\ff
with $p^{||}= -{\ttau_{ai}}{\tilde E}^{ai}/(8\pi{G})$ and
$\tilde{p}_\perp^{ai} = -H_{bj}^{\ \ \ ai}{\tilde E}^{bj}/(8 \pi G)$.

We may also write down directly the momentum conjugate to the
Chern-Simons invariant $\tau_{CS} = \int_\Sigma Y_{CS}$.  It is
\f
p_{CS} \equiv {1 \over {\cal B}}  \int_\Sigma \tilde{b}p^{||} ,
\ff
where ${\cal B} \equiv \int_\Sigma \tilde{b}$.
It may be checked directly that $\{ \tau_{CS} , p_{CS} \} =1$.

We may now
proceed to construct the Hamiltonian conjugate to $\tau_{CS}$. We
plug (29)
into the Hamiltonian constraint (14), to find,
\begin{eqnarray}
0 = {\cal H} &=&(8\pi G)^2 [ (p^{||})^2 \tilde{b}^2 -
\tilde{p}_\perp^{ai}
\tilde{p}_\perp^{bj} \tilde{b} \tH_{ai,bj}
\nonumber \\
&&-(16\pi \lambda)\left( (p^{||})^3 {\tilde b}^2 +
{1\over 2}p^{||}{\tilde b} \tH_{ai,bj} \tilde{p}_\perp^{ai}
\tilde{p}_\perp^{bj}+ det(\tilde{p}_\perp^{ck})
\right)]
\end{eqnarray}
To proceed we must consider separately the cases
in which there is or is not a cosmological constant.
For the remainder of this section we set $\lambda$ to
zero, postponing to the next two sections the case of
nonvanishing cosmological constant.
With no $\lambda$ terms,
we may solve (31) to find a hamiltonian that generates
evolution in $\tau_{CS}$.  First, we find directly
that
\f
p^{||}(x) =\sqrt{\tilde{b}^{-1}\tH_{ai,bj}(x)
\tilde{p}^{ai}_\perp (x) \tilde{p}^{bj}_\perp (x)}
\ff
To find the Hamiltonian conjugate to $\tau_{CS}$ we
must integrate.  We find
\f
p_{CS}
= {1 \over {\cal B}}  \int_\Sigma
\sqrt{\tilde{b}\tH_{ai,bj} \tilde{p}_\perp^{ai} \tilde{p}_\perp^{bj} }
\equiv h_{CS}
\ff
This suggests that a quantization could be developed along the
lines of \cite{carlolee-ham} in which a quantum state,
$\Psi$ of the gravitational field,
expressed either in the connection or the loop representation,
evolves according to
\f
\imath {\partial \Psi\over \partial \tau_{CS}} =
\hat{h}_{CS} \Psi
\ff
We may note that as we have defined the
Hamiltonian without breaking spatial
diffeomorphism invariance,
it may be possible to implement the corresponding time
evolution equation quantum
mechanically as an operator on diffeomorphism invariant states,
using the techniques developed in \cite{carlolee-ham}.
Furthermore, we may note that
as  $\tH_{ai,bj} $ is positive definite,
the square root in (33), and hence the Hamiltonian,
may be defined over all regions
of the configuration space  ${\cal C}_{diff}$ on which
$\tilde{b}$ is also positive definite everywhere on $\Sigma$.
The only difficulties with the definition of this Hamiltonian
then  come through the possibility of vanishing or
complex $\tilde{b}$.   This is a better situation than the
Hamiltonian obtained in \cite{carlolee-ham} which is
the square root of an expression that is not positive
definite at generic points
on the configuration space.  This is at least one
advantage of using a degree of freedom of the gravitational
field, rather than that of a matter field, as a clock. Eqn.(34) suggests
that for $\lambda =0$, there may be ``stationary states" of the
form $\Psi[A] = e^{c\int_{\Sigma}Y_{CS}}$$\chi_{\gamma}[A^\perp]$,
with $c$ being dimensionless
constants which correspond to the ``spectrum" of $h_{CS}$.

Finally, we may note that
an even simpler form for the Hamiltonian is obtained
if we scale the time coordinate by ${\cal B}$, so that we
evolve the state not in $\tau_{CS}$, but in
\f
\tau^\prime = {\tau_{CS} \over {\cal B}}
\ff
The hamiltonian should then be proportional to
 $p^\prime= \int_\Sigma \tilde{b} p_{CS}$, which is simply,
\f
p^\prime
= \int_\Sigma
\sqrt{\tilde{b}\tH_{ai,bj} \tilde{p}_\perp^{ai} \tilde{p}_\perp^{bj} }.
\ff

Now we turn, in the next two sections, to the case of finite
cosmological constant.

\section{The physical interpretation of the Chern-Simons state}

The case of finite cosmological constant, is more difficult,
because of the term in (31)
proportional to the cube of $p^{||}$.  This means that
the quantum
hamiltonian constraint equation contains third  time derivatives
in the apparent local time variable $\tilde{A}^{||}$.
The presense of these
third time derivatives poses a difficulty for the interpretation
of the dynamics of the theory, as they could be
indications that the theory has  instabilities or
runaway solutions, of the kind that infest the Lorentz-Dirac
formulation of the relativistic electrodynamics of point
particles.  We may note, in this connection,
that working in perturbation
theory, Tsamis and Woodard have found evidence that
quantum general relativity with a finite cosmological
constant is infrared unstable\cite{richardnick-cc}.   Thus,
we should be cautious about the interpretation of a quantum
theory for nonvanishing $\lambda$.

On the other hand, it may be that in spite of these
cubic terms, the quantum theory with a cosmological constant
is still well defined, at least for sufficiently small $\lambda$.
We may note that exponential growth
is a property of the cosmological solutions associated
with the presense of a cosmological constant, thus third time
derivatives may be necesssary if the quantum theory with
a cosmological constant is to have a good semiclassical limit.

But perhaps the best evidence that the theory may be
sensible in the presence of a cosmological constant is that
we know that in that case there is an exact quantum
state of the theory,
which is the Chern-Simons state discovered by Kodama\cite{kodama}.
This is an exact solution to all the constraints of quantum gravity
in the connection representation
for $\lambda \neq 0$, which is
given by\footnote{We may note that the
Chern-Simons state may be multiplied by
any  topological invariant, $I$, of $A$ on ${\Sigma}$\cite{cso}.  We
may note that, as $I$ may be chosen arbitrarily without
affecting the demonstration that  $\psi_{CS}$ solves
the constraints, there could actually be a number of such
exact Chern-Simons states which depend on the topology of
$\Sigma$.
We do not here develop this very interesting fact.},
\f
\psi_{CS} [A] \equiv
e^{{3 \over 16{\pi}\lambda} \int_{\Sigma} Y_{CS} [A]}
\ff
 This state has been much studied in quantum gravity, and it
is known that a class of states with very interesting
properties can be constructed by transforming it to
the loop representation\cite{BGP-cs}.

The question of whether the quantum theory with a finite
$\lambda$ can be sensible then depends to some extent on
the interpretation of this state.   While we are not
able to settle this question here, we are able to discover a
significant piece of evidence that the Chern-Simons state might,
at least for sufficiently small $\lambda$, be interpreted as
the ground state of the theory.  This evidence is
gotten by coupling the theory to matter,
and then studying the behavior of perturbations of this state
involving the matter degrees of freedom.
We find that these excitations
behave, to
leading order in $\lambda$, exactly  like excitations of a quantum
field on the background of a DeSitter spacetime.  Furthermore,
we find that in this description the natural time coordinate
in which these matter states evolve is the Chern-Simons time.

We then consider an excitation of a
matter field, $\phi$, defined by
a state,
\f
\Psi [A, \phi ] = \psi_{CS} [A] \chi [A, \phi ]
\ff
For concreteness, we will take the matter field to be a single
massless free scalar field, although the results are
independent of the choice.
We then apply the Hamiltonian constraint to this state, using a
regularization and an ordering
in which $\psi_{CS}[A]$ is an exact
solution\cite{kodama, cso, BGP-cs} and find,
using $\tilde{E}^{ai} (x) = - (8\pi G)  \delta / \delta A_{ai}(x)$,
that,
\begin{eqnarray}
0&&=\int_\Sigma {\tN} \left \{  -{{3G} \over {2 \lambda} }
\tepsilon_{abc}\epsilon_{ijk}
\tilde{B}^{ck} \tilde{B}^{bj} {\delta \over \delta A_{ai} }
+ {\cal H}_{matter}[ \tilde{E}^{ai}= -{{3G} \over \lambda }
\tilde{B}^{ai} ]
\right \} \chi [A, \phi ]
\nonumber \\
&&+\int_\Sigma{\tN} \left \{(\partial_a \phi )(\partial_b \phi )
({{3G} \over \lambda } \tilde{B}^{(a}_i )(8\pi G){\delta \over \delta
A_{b)i} } +(\partial_a \phi )(\partial_b \phi )
{(8\pi G)}^2 {\delta \over \delta A_{a}^i }
{\delta \over \delta A_{bi} }
\right \} \chi [A, \phi ]
\nonumber \\
&&-G{\lambda \over 6}\int_\Sigma{\tN}{(8\pi)}^2
\epsilon^{ijk} \tepsilon_{abc}
{\delta \over \delta A_{ai}}{\delta \over \delta A_{bj}}
{\delta \over \delta A_{ck}} \chi [A, \phi ] +...
\end{eqnarray}

To analyze the meaning of this equation, let us assume that
$\chi [A,\phi ]$ is peaked around self-dual spacetimes,
and has only a slow dependence on $A_{ai}$ compared to the
leading exponential term in the Chern-Simons state.
In this case, the terms in the second and
third lines of (39) are of lower order compared to the first.
Using (22), we then have, to leading order
\begin{eqnarray}
{{3G} \over {2 \lambda} } \int_\Sigma {\tN}
\tilde{b}^2 \ttau_{ai} {\delta \chi [A, \phi ] \over \delta A_{ai} }
&=&
 {{3G} \over {2 \lambda} } \int_\Sigma {\tN}
\tilde{b}^2 {\delta \chi [A, \phi ] \over \delta A^{||} }
\nonumber \\
&=&\int_\Sigma {\tN}
{\cal H}_{matter}[ \tilde{E}^{ai}= -{{3G} \over \lambda }
\tilde{B}^{ai} ]
\chi [A, \phi ]
\end{eqnarray}
Now, let us assume that the dependence on $\chi [A,\phi ]$
on $A^{||}$ is holomorphic, following the usual analogy between
the Ashtekar connection and the Bargmann quantization of the
harmonic oscillator.  Then, we may write
\f
{\delta \chi \over \delta \tilde{A}^{||}(x) }
= \imath {\delta \chi \over \delta Im \tilde{A}^{||}(x)}
\ff
It is also possible to show that
in the limit of small $\lambda$ with ${\tilde E}^{ai}$ approaching
$\delta^{ai}$, on the L.H.S. of the Schroedinger equation,
\f
{\delta \chi \over \delta Im \tilde{A}^{||}(x)}=
{\delta \chi \over \delta Im Y(x)}    + O (\lambda )
=  {{3G} \over { \lambda} }  {\delta \chi \over \delta \tilde{K}}
+ O(\lambda )
\ff
where $\tilde{K}=\sqrt{q}K$ is the densitized trace of the
extrinsic curvature, and we have used (3).

Thus we have a local Schwinger-Tommonoga equation of the form
\f
 {\imath \over 2} \left ( {{3G} \over { \lambda} } \right )^2
{\delta \chi \over \delta \tilde{K}}=
\tilde{b}^{-2}{\cal H}_{matter}[ \tilde{E}^{ai}= -{{3G} \over \lambda }
\tilde{B}^{ai} ]
\chi [A, \phi ]  + O(\lambda )
\ff

Alternatively, we can find a single Schroedinger equation that
governs the propagation of the state in the Chern-Simons  time.
To find this, we integrate (40), with $\tN = \tilde{b}^{-1}$, to find,
again using (3),
\f
\imath {\delta \chi \over \delta \tau_{ICS}} =
{18 G^5 \over \lambda^2 V(q) }
\int_\Sigma \sqrt{q} H_{matter} \chi
\ff
where $H_{matter}=\tilde{q}^{-1}{\cal H}_{matter}$ is
the undensitized matter hamiltonian and
$V(q) =\int_\Sigma \sqrt{q} \approx
(3G/ \lambda )^{3/2}\int_\Sigma  \tilde{b} $ is the spatial volume
of the universe.

These equations tell us that for small
cosmological constant,  the leading term of the Hamiltonian
constraint can be interpreted as the quantum field theory for the
matter
fields evolving with respect to the Chern-Simons
time $  \tau_{ICS}$ on a classical background manifold which obeys
$
\tilde{E}^{ai} = -{3G \over { \lambda}}B^{ai}$.
But the reality conditions\cite{abhay-reviews} then
imply that the background must be a DeSitter spacetime.  Thus, in
the
limit of small $\lambda$, and excitations energies small in Planck
units,
the excitations of the exact state may be interpeted as
describing quantum fields evolving on a background DeSitter
spacetime.
Further, the natural time coordinate on that DeSitter
spacetime that emerges from this limit of the
Hamiltonian constraint is precisely
$ \tau_{ICS}$.

In closing this section,
we note that  one further comment may be made
about the hypothesis that the Chern-Simons state describes
the vacuum of the theory.
We may note that it is indeed remarkable that a state that is
an exact eigenstate of momentum such as
is $\psi_{CS} [A]$ can have an
interpretation of an exact vacuum of
a quantum field theory.  It is indeed for this
reason that there has been a suspicion that the Chern-Simons
state is unphysical, for such is certainly the case for the
corresponding solution of Yang-Mills theory.   The difference
between these two cases lies in the fact that in quantum gravity
the inner product reflects the dynamics of the theory, and must
be imposed on the space of solutions to the theory.
If the Chern-Simons invariant is time, and if the inner product is
taken
at fixed time, as we will shortly propose,
then there is a possibility that this state may
be
normalizable.  This, and the
fact that it is only the inner product that will impose the
reality conditions, means that
a state may be simultaneously an exact solution to the constraints
in an ordering that is, as we have shown here, consistent with the
semiclassical limit {\it and} a Hamiltonian-Jacobi function for
a sector of solutions of the theory.

Further, we note that because the Chern-Simons state is
simultaneously a semiclassical state and an exact state,
the semiclassical expansion is cleaner than in the conventional
formulation\cite{banks-etc},
so that there is no DeWitt-Morette-Van Vleck determinant.
The Chern-Simons state must then already contain information about
the linearized virtual excitations of the gravitational field.

This means that the Chern-Simons state
corresponds to a universe in which only the left-handed curvature
is virtually excited.  This is sensible
because, as the
reality conditions are imposed by the inner product, it need
not be satisfied by the virtual quantum excitations.  This
suggests that the Chern-Simons state describes a
vacuum at the Planck scale that is
naturally chiral, corresponding to a condenstate of left-handed
{\it nonlinear}\cite{roger-nonlinear}
gravitons, while at the same time reproducing the
physics of  classical general
relativity in the semiclassical limit.  This suggest that quantum
gravity may naturally have $CP$ violating effects at short
distances.  Further evidence for this conclusion is in \cite{cp-on-cp}.

\section{Gravitational perturbations with respect to the conformally
self-dual sector}

In the previous sector we saw that, in the presence of a cosmological
constant, it is possible to make the hypothesis that the ground state
of quantum gravity is given by the Chern-Simons state.  To further
investigate this hypothesis, it would be necessary to study the
higher order corrections to the $WKB$ limit we have just discussed.
While we are not yet
able to do this, we make some preliminary remarks
in this section.

As we noted, the Chern-Simons state can
be understood as being the exponential of
a Hamilton-Jacobi function for the
self-dual sector.  Thus, if we want to understand how to study
perturbations of the quantum theory from this vacuum, we must
first understand how to describe
perturbations
away from self-dual sector in the classical theory.  In
this section we study this question by studying the perturbations
of the classical Hamiltonian constraint away from the self-dual
sector.
To do this, it is important to observe that there are two
dimensionless
parameters of interest,
\f
{\tilde q}^{ai} \equiv {\tilde E}^{ai} + {{3G}\over \lambda}{\tilde
B}^{ai}
\ff
and $\lambda$, the cosmological constant.
${\tilde q}^{ai}$ vanishes for conformally self-dual
solutions, such as the DeSitter solution, and thus describes
fluctuations away from the conformally self-dual sector.
Thus, to expand around the DeSitter solution, for
universes large relative to the Planck scale, we may expand
simultaneously
in small $\tilde{q}^{ai}$ and small $\lambda$.

We may then proceed by separating
the terms of different orders of  $\lambda$ and
${\tilde q}^{ai} $ in the action of the
classical Hamiltonian constraint.    We
consider first, the case with matter, in the form of
a scalar field, as in the previous section.  We may note that
the classical hamiltonian constraint can be written in terms of the
fluctuations ${\tilde q}^{ai}$ as
\begin{eqnarray}
0 &=&\tepsilon_{abc} \epsilon_{ijk}[{\lambda \over
G}{\tilde q}^{ai}{\tilde q}^{bj}{\tilde q}^{ck} -
6{\tilde B}^{ai}{\tilde q}^{bj}{\tilde q}^{ck} +
{{9G{\tilde B}^{ai}{\tilde B}^{bj}}\over \lambda}{\tilde q}^{ck}]
\nonumber \\
&&-(16{\pi} G)[{1 \over 2} {\tilde \pi}^2 + {1\over 2}({\tilde
q}^{ai}{\tilde q}^{bi} -
{{6G}\over \lambda}{\tilde B}^{(ai}{\tilde q}^{b)i} + {{9G^2}\over
\lambda^2}
{\tilde B}^{ai}{\tilde B}^{bi})\partial_a\phi\partial_b \phi]
\end{eqnarray}

 First, we may expect that
for small ${\tilde q}^{ai}$,
the first and last terms in the matter hamiltonian are dominant
(${G{\tilde B}}\over \lambda$ is of order one).
For small ${\tilde q}^{ai}$, the dominant contribution to the pure
gravity hamiltonian comes from the last term which is first order in
${\tilde q}^{ai}$. Therefore, to leading order, in the
presence of matter we have
precisely
\f
{9 G \over \lambda }\tepsilon_{abc} \epsilon_{ijk}
{\tilde B}^{ai}{\tilde B}^{bj}{\tilde q}^{ck}
=(16{\pi}G)[{1\over 2}{\tilde \pi}^2 + ({{9G^2}\over{2\lambda^2}})
{\tilde B}^{ai}{\tilde B}^{bi} \partial_a\phi\partial_b \phi ].
\ff

But this, evaluated at
$\tilde{E}^{ai} \approx -{3G \over {2 \lambda}}{\tilde \epsilon}^{abc}F_{bc}^i
$, corresponds to the semiclassical
limit of the Wheeler-DeWitt equation,
as we found in the last section.  It tells us how the leading
order deviations from the conformally self-dual sector are
driven by the matter fields.

What happens in the absense of matter?  Here the circumstance
is somewhat different, as the leading gravitational term cannot
be balanced against the matter hamiltonian.  Instead, we may
seek solutions in which the different terms in the gravitational
perturbations are balanced against each other.

To find such solutions, let us note that
for small $\lambda$
and ${\tilde q}^{ai}$,
we may neglect the term which is third order in
${\tilde q}^{ai}$ in (46).
The resultant hamiltonian constraint up to terms quadratic in
${\tilde q}^{ai}$ is then of the form
\f
q^2 -{9 \over 2}q +{{3\lambda^2}\over {2 G^2{\tilde b}}}
\tH_{ai, bj}{\tilde q}^{ai}_{\perp}{\tilde q}^{bj}_{\perp} = 0
\ff
where $q \equiv {3\lambda\over G}{\tilde q}^{ai}\ttau_{ai}$ and
${\tilde q}_\perp^{ai}\equiv
{\tilde q}^{ai}-{G\over 3\lambda}{\tilde \tau}^{ai}q$.
Note that $\ttau_{ai}{\tilde q}^{ai}_{\perp} = 0$, and $q$ is conjugate
to
$\int_{\Sigma}Y_{CS}$.
We may then solve this to find
\begin{eqnarray}
q &\equiv& {\cal H}_{\perp} ({\tilde B}, {\tilde q}_{\perp} )
\nonumber\\
&=& {9 \over 2}
\left( 1 - \sqrt{ 1-{{8\lambda^2}\over {27 G^2 {\tilde
b}}}
{\tH_{ai, bj}{\tilde q}^{ai}_{\perp}{\tilde q}^{bj}_{\perp}}}\right)
\end{eqnarray}
It is interesting to note that for small perturbations, this has the
form of
\f
q = {\cal H}_{\perp} = {{2\lambda^2}\over {3 G^2 {\tilde b}}}
\tH_{ai, bj}{\tilde q}^{ai}_{\perp}{\tilde q}^{bj}_{\perp} +
O({\tilde q}_{\perp}^4)
\ff
This shows us that, for small $\lambda$,
the gravitational perturbations of conformally
self-dual solutions are stable, at least for the
Euclidean sector in which $\tilde{b}^2 >0$.  This is
a necessary, but of course, not sufficient, condition for the
demonstration of the stability of the theory for
nonvanishing cosmological constant.

\section{Proposal for an inner product}

One of the most difficult problems in non-perturbative quantum
gravity is the construction of the inner product.  While this, also,
is a question that we cannot completely settle here, we are able
to note that the hypothesis that the Chern-Simons invariant gives
us a natural notion of time on the configuration space of general
relativity leads to a suggestion for a form of the inner product.

This suggestion is, in principle, very straightforward: if there is
a natural time coordinate on the configuration space, then the
inner product may be chosen by integrating on a constant
time surface of the configuration space.  A related
suggestion has recently been put forward by
Moncrief\cite{vince-ip},
who suggested that an algebra of physical observables can be
constructed from the values of physical fields on the
$K=0$ surface of each spacetime.  More particularly, he noted that,
at least for a large region of the physical phase space which
corresponds to solutions that have
$K=0$ surfaces, a set of physical observables is given by a
complete set of spatially diffeomorphism invariant functions
of the fields on that surface.  These functions have an algebra
which is easy to compute, for one can simply evaluate their Poisson
brackets, at each solution in the physical phase space, in terms of
the Poisson brackets of the fields on the
 $K=0$ surface of that spacetime.

This proposal has been explored in a study of a
nonperturbative quantization
of the Bianchi IX cosmologies\cite{sethlee-canonical}.
Here we would like to show that a rather suggestive form for
the inner product emerges if we take Moncrief's suggestion
seriously, but with two modifications.  First, we take the
special surface to be identified on the configuration space
rather than on each spacetime.  This is in line with our
philosophy that the configuration space, rather than the spacetime,
is the proper arena for the quantum theory.   Second, we take
the surface of initial time on which the inner product will be
defined to be the surface
in the configuration space where the
imaginary part of the Chern-Simons
invariant vanishes.  Of course, by (3) this is  closely related
to the condition that $K$ vanishes.
However, taking the condition in terms of an invariant on
the configuration space means
that a gauge condition must be fixed to identify which surfaces
in a particular solution correspond to the $ Im \int Y=0$
surface of configuration space.  This allows us to write the
inner product in a more gauge-invariant form.

Our proposal is then to consider an inner product
of the form
\f
<\Psi^\prime | \Psi > = \int \prod \left ( (dA)(d\bar{A})
\delta [ \tau_{ICS}(A)]   \right  )
\bar{\Psi}^\prime (A) \Delta [\bar{A}, A]  \Psi (A)
\ff
This is not a complete definition, for
the expression $ \Delta [\bar{A}, A] $ must be chosen so that
the physical Hamiltonian, defined appropriately on such states is
Hermitian,
and will also contain gauge fixing factors.
The form of $ \Delta [\bar{A}, A] $ that satisfies this condition
has not yet been found and involves difficult issues of
ordering and regularization.  However, we may notice that,
even without a complete specification of the inner product,
we may see that this general form leads
to
a suggestive formal expressions for the inner products
of certain states.  For if we take seriously the suggestion
that the Chern-Simons state is
be the vacuum of the theory, we may consider
modified loop states of the form
\f
\Psi_\gamma [A] \equiv e^{{3 \over {16\pi \lambda}}
\int_{\Sigma}Y_{CS}[A]} T_{\gamma}[ A]
\ff
where $T_{\gamma}[ A]$ is the usual product of the traces of
holonomies of loops.
The functional integral () defining the inner product
may for such states be expressed
in terms of a functional integral involving the
$SO(3)$ connection $\Gamma_{ai}$ and the extrinsic curvature,
$K_{ai}$.
\begin{eqnarray}
<\Psi^\prime | \Psi > &=&
 \int \prod \left ( (d\Gamma_{ai} )(dK_{ai})
\delta [Im \int Y(\Gamma, K)]
\right  )
\nonumber \\
&&\times e^{{6 \over {16\pi \lambda} }
\int_{\Sigma}(Y_{CS}[\Gamma]
-{\tilde \epsilon}^{abc} K_a^i D^\Gamma_b K_{ci}) }
\nonumber \\
&&\times
\bar{T}^{\prime}_{\gamma} [\Gamma , K] \Delta [A=\Gamma +
\imath K]
T_{\gamma}[\Gamma , K]
\end{eqnarray}

This inner product thus defines a field theory
for a multiplet of vector fields, $K_{ai}$ interacting in a background
given by the $SO(3)$ gauge field $\Gamma_{ai}$.  As
the exponential in (53) is quadratic in
$K_{ai}$ one may develop the integral over $K_{ai}$
perturbatively by a power series expansion in $K_{ai}$.  Meanwhile,
as $T_{\gamma}[A]$ is a product of holonomies of
loops,
the integral over $\Gamma$ might be defined by using Witten's
discovery that
the Chern-Simons path integral yields the Jones polynomial of knot
theory\cite{witten-cs}.   If this can be done, it will
support the conjecture that $\psi_{CS}[A]$ is indeed the physical
ground state of the theory, by giving meaning to its gravitational
excitations of the form of (52).

In closing this section,
we may note that in Section 5 we found evidence that
the imaginary part of the Chern-Simons invariant is playing
the role of time only in the semiclassical limit.  The evidence
we have for the role of the Chern-Simons
invariant that is beyond the semiclassical limit
comes from the role of the
real Chern-Simons invariant in the Euclideanized sector.
Niether completely justifies the step we are taking here,
which is to hypothesize that $\tau_{ICS}$ plays the
role of time
for the full, non-perturbative theory.  This proposal can
only be confirmed if it leads in fact to physically significant results.

\section{The KMS condition for nonperturbative quantum
gravity}

In this section we would like to discuss a
final remarkable consequence
of the choice of the Chern-Simons invariant as the time function on
the configuration space.  This is that the
full
nonperturbative quantum theory may
be intrinsically thermal, in that all expectation values of physical
observables satisfy a $KMS$ condition\cite{kms}.

To see why, note that if we study the real section of
${\cal C}_{kin}$ corresponding to Euclidean
spacetimes,
$\tau_{CS}$ becomes a periodic coordinate
because under large gauge transformations of winding number
$n$, the
Chern-Simons form is not invariant but transforms as
$\int Y_{CS} \rightarrow \int Y_{CS} + 8{\pi^2}n $ \cite{cs}.
This means that the real section
of ${\cal C}_{diff}$ has the topology of
$S^1 \times {\cal S}$, where ${\cal S} $ is the slice of
${\cal C}_{diff}$ defined by $\int Y_{CS}=0$.
This has a direct consequence, which is the following.

Let us assume that a complete quantum theory of gravity has been
defined as a quantum theory of motion on the configuration
space ${\cal C}_{diffeo}$.
This should allows us to calculate
expectation values of the form
\f
<\Psi |\hat{\cal O}_1 \hat{\cal O}_2 ... |\Psi >
\ff
where the ${\cal O}_i$ are physical, and hence gauge
invariant, observables, the $|\Psi>$'s are physical
states and  the expectation value
is defined in terms of a physical inner product.
Now, the problem of how to construct physical observables
that measure evolution in quantum gravity,
is well understood, at least in principle
(see, for example, \cite{carlo-time}).
What is clear is that such observables describe correlations between
certain degrees of freedom, which we take as representing a
measure of time, and other degrees of freedom of the theory.
It then follows from this that given
any choice, $\tau$ for a measure of time on the physical
configuration space it will be the case that for every
diffeomorphism invariant function $O[A,E]$ on the
phase space
there will be one parameter family of physical observables
in the classical theory,
${\cal O}(t)$, that measure the value of $O[A,E]$
averaged over the configurations for which $t=\tau$.
For example, such a function might correspond to the measurement
of some spatially diffeomorphism invariant function, $O$, of
$A_{ai}$ and $\tilde{E}^{ai}$ on slices that satisfy a certain
gauge condition, such as maximal slicing.  The definition of
the observable, as a function on the physical phase space or,
equivalently, on the space of solutions, would be to find that slice
in the slicing of each solution for which $\tau=t$, and then
measure $O$ on that slice.

We may then assume that in a succesful quantum theory of gravity
there will be one parameter families of quantum observables,
$\hat{\cal O}(t)$, corresponding to some subset of
these classical observables.
It is natural to
assume also that in such a theory there will be a functional
integral representation of physical expectation values in
terms of paths on the diffeomorphism invariant configuration
space ${\cal C}_{diffeo}$.  In this case,  then we may expect that
there will be a measure $\mu (A)$ on ${\cal C}_{diff}$,
such that, at
least for a preferred vacuum state $<0|$,
\f
G^{O}(t) \equiv <0 |\hat{\cal O}_1(t) \hat{\cal O}_2 (0) |0> =
\int_{{\cal C}_{diff}}d\mu(A) {\cal O}_1(t) {\cal O}_2 (0)
\ff
Now, given that this will exist, it is natural also to
imagine that in Euclidean quantum gravity the expectation values are
to be measured in terms of the path integral (55) over Euclidean
configurations. If we want to interpret these
expectation values in terms of a Euclideanized time, we
have to follow the same procedures as in the physical theory,
which is to find a good time coordinate $\tau_{Euc}$
on the physical
configuration space, ${\cal C}_{diff}$ and construct
operators that measure correlations between this
degree of freedom and other degrees of freedom of the
theory.

Now,  we have found in the preceeding
sections that there is a preferred choice for the Euclideanized
time of the theory, which is $\tau_{CS}$, the Chern-Simons
invariant.  It is therefore natural to investigate the behavior
of the Euclideanized quantum field theory when the observables
are parameterized by this notion of time.
It is very interersting to note that
it follows immediately, from the periodicity of
the configuration space for Euclidean spacetimes ${\cal C}_{diffeo}$ just
mentioned that \f
G^{O}_{Euc}(\tau_{CS})=G^{O}_{Euc}( \tau_{CS} + 8 \pi^2n )
\ff

Thus, from the geometry of the diffeomorphism invariant
configuration space, it must follow that, to the extent that
a quantum theory of gravity exists along the conventional lines
we have assumed here, physical expectation values in the theory
must satisfy the $KMS$ condition, when correlation functions
expressed in terms of $\tau_{CS}$ are measured.
This means that there is a sense in which quantum general relativity must
be an intrinsically thermal theory.

We may note that it may also be the case that the
Euclidean amplitude
$G^O_{Euc}[\tau_{CS}]$ is the analytic continuation of the
physical amplitude $G^O_[\tau_{ICS}]$.  However, given the
well known difficulties about Euclideanization in quantum gravity,
this should not be assumed.  It is then important to emphasize that
it is not necessary that this be the case for it still to follow
that the Euclideanized quantum theory satisfied a $KMS$
condition when expressed in terms of a natural notion of
time for the Euclideanized theory.

Of course, this is a formal argument.  However, it is interesting
to note that it can be directly confirmed in the semiclassical
limit.  To show this we must begin by asking an important
question:  What is the temperature associated
with the periodicity we have found?  The answer is
that, because the period is
a dimensionless parameter, there is actually no dimensional
temperature associated with the theory.  Instead,
a particular temperature can arise from the periodicity of
the Euclidean signature Chern-Simons invariant only for the case of
a semiclassical state.  In that case the amplitudes are dominated
by a particular periodic trajectory on the configuration space,
corresponding to a Euclidean signature spacetime that is periodic in
Euclidean time.  Then the metric of this spacetime can provide a
dimensional measure of the periodicity, which then provides
a temperature for the fluctuations around it.

For example, to
find the temperature associated with the
Chern-Simons state (37),
which we conjecture to be the vacuum,
we use the fact established above
that this state can be understood in
the semiclassical limit as describing fluctuations around
DeSitter spacetime.  It then must follow that
the periodicity of the real section of ${\cal C}_{diff}$
must corresponds to the known
periodicity of the Euclidean DeSitter
spacetime\cite{GHaw}.
In fact, we can use the invariance
of the theory under large gauge transformations to discover
the periodicity of Euclidean DeSitter spacetime.
For  constant-$t$ foliations of DeSitter spacetime,
the relation between the Chern-Simons
invariant and the spacetime coordinate $t$ is
\f
{\partial\tau_{CS} \over \partial t} =
\int_{\Sigma}dx{\tilde B}^{ai} \{ A_{ai} , {\tN}{\cal H}\}_{P.B.}
=
\int_{\Sigma} d^3 x {\tN} \tepsilon_{abc}\epsilon_{ijk}
{\tilde B}^{ai}{\tilde E}^{bj}({\tilde B}^{ck} +
{\lambda \over 2G}{\tilde E}^{ck})
\ff
In arriving at expression (57), which is the {\it general relation between
the variable t in a classical t-foliated spacetime and the Chern-
Simons functional}, we have used ${\partial \tau_{CS}\over \partial
A_{ai}} = {\tilde B}^{ai}$. The DeSitter metric may be written as
\f
ds^2_{DeSitter} =
 (1-{\lambda r^2\over 3G})dt^2
+ {1 \over (1-{\lambda r^2\over 3G})}dr^2 + r^2 d{\theta}^2 +
r^2sin^2{\theta}d{\phi}^2
\ff
On substituting $\sqrt {det(\tilde E)}{\tN}$ for the lapse function,
$\sqrt{1-{\lambda r^2\over 3G}}$, and integrating over
$ 0 \leq \phi < 2\pi, 0\leq \theta <\pi, 0 \leq r
< \sqrt{3G \over \lambda}$,
we arrive at
\f
{d \tau_{CS} [A_{DeSitter}(t)]) \over dt} = - 4\pi
\sqrt{\lambda \over 3G}
\ff
However, by integrating this we find
that the invariance of the theory under
large gauge transformations requires that for
the Euclidean DeSitter manifold, $t$ must be
periodic with period $2\pi \sqrt{3G/ \lambda}$.
This also gives us the scaling between the
dimensionless period of $\tau_{CS}$ and a physical temperature,
which then implies
that the temperature associated with the Chern-Simons state
is indeed,
\f
T_{cosmological} = {1 \over 2\pi} \sqrt{\lambda / 3G}  .
\ff
This coincides with the periodicity and the Hawking temperature
found by
demanding that the metric of the Euclidean DeSitter solution
have no conical singularity\cite{GHaw}.

In the case of zero cosmological constant, an analogous calculation for
the
Euclidean Schwarzschild solution with metric
\f
ds^2_{Schwarzschild} =
 (1-{2Gm\over r})dt^2
+ {1 \over (1-{2Gm\over r})}dr^2 + r^2 d{\theta}^2 +
r^2sin^2{\theta}d{\phi}^2
\ff
gives a similar correspondence
\f
{{d\tau_{CS}[A_{Schwarzschild}(t)]}\over {dt}} = \pi/Gm
\ff
between the periodicity of the Euclidean Chern-Simons
term of the Ashtekar-Sen connection and the periodicity of the
Euclidean
time coordinate with period $8\pi{Gm}$.

As a result, we may conclude that the use of the connection
configuration space, together with the conjecture that
the Chern-Simons invariant is a natural time coordinate,
gives us a setting in which the connection between periodicity
in the Euclidean sector and thermal states may be extended from
the semiclassical theory to the full nonperturbative theory.
We thus conjecture that quantum gravity may be an intrinsically
thermal theory,  not only
at the semiclassical level\cite{GHaw,GHP} but also at the full
non-perturbative level.

As a final note to this section, we may observe that the idea
that the absolute distinction between quantum and thermal
fluctuations might break down in quantum gravity has
been rasied before\cite{ls-thermal}.   Furthermore,
for rather different reasons,   Connes and Rovelli
have found independent reasons to
conjecture that general states of quantum gravity satisfy
a $KMS$ condition\cite{alaincarlo}.  In their ``Thermal
Time Hypothesis" they propose that a natural notion of time
may be derived from a large class of quantum states of
a diffeomorphism invariant quantum field theory such as
general relativity.  It is worth exploring whether the
Chern-Simons time may be a realization of their hypothesis.

\section{Conclusions}

The diffeomorphism invariance of classical general relativity
means that the coordinates on any given spacetime manifold have
no physical significance.  Instead, time evolution is to be
described, in the classical theory, in terms of correlations between
different degrees of freedom of the theory.  As any of the infinite
degrees of freedom may, at least locally, be taken as a clock,
classical general relativity then allows an infinitude of possible
descriptions of a unvierse evolving in time.

The problem in time in quantum gravity, especially in the
cosmological case, then has two aspects.  First, is it possible in
the quantum theory, as it is in the classical theory, to describe
time evolution in terms of an infinite variety of possible clocks?
Second, is it practical to do so?  That is, even if it is possible to
formally define the theory in such a way as to accomodate
evolution by any possible clock, might it be more convenient to
define the theory in a simpler way so that evolution with respect
to some particular clock is described.  Whatever the answer to
the first question, which may be considered to be the deep problem
of time, it still may be much more convenient to define the theory
with respect to a particular notion of evolution, with respect to
which the equations may be particularly simple, rather than in
a form allowing interpretations in terms of all possible clocks.

We certainly did not answer the deep question of time here.  But
what was accomplished was to show that the classical configuation
space of general relativity has on it a natural function,
the Chern-Simons invariant, which may
serve as a useful time coordinate for the theory.  Furthermore, we
saw that this function endows the
configuration spaces ${\cal C}_{kin}$ and ${\cal C}_{diff}$ with
certain structures which are very convenient for understanding
the dynamics of general relativity. This structure, which
we described in Sections 2 and 3 represents
a certain special kind of geometry in which the metric and geodesic
structure of an infinite dimensional space are defined in terms of
a simple
function.  As we saw in various guises in the course
of this paper, it
gives us a framework for analyzing dynamical questions
in general relativity, which
may be of use for understanding a variety of features of the dynamics of
classical
and quantum general relativity.

At the level of the classical theory, we have found that a study of
the role that the Chern-Simons invariant plays in the geometry
of the configuration space of general relativity
leads to the following results:

i)  A splitting of the physical degrees of freedom into gauge,
physical, and time like degrees of freedom (Section 2).

ii)  The discovery of simple forms for the action principle, as
a geodesic principle on the configuration space ${\cal C}_{diff}$
(Section 3).

iii)  The existence of a canonical transformation that separates
the timelike and physical degrees of freedom, and that leads
to particularly simple forms for Hamiltonians that evolve
the physical degrees of freedom of the gravitational field
in terms of time parameters related to the Chern-Simons
invariant (Sections 4, for vanishing cosmological
constant and 6 for nonvanishing $\lambda$).

We may note that all of these results have been written here
for the Euclidean signature case, which is the simplest when
dealing with the Ashtekar formulation.  If we want to discuss
the Minkowskian case we must impose reality conditions.
As these
mix coordinates and momenta, they are not easily
expressed in terms
of the configuration space variables.    The effect of the reality
conditions on these results, in both the classical and quantum
theory, is not yet well explored.  We may note
only that the expectation
that quantum states will be, under a certain definition of
complex structure, holomorphic in the connection, suggests that these
and other results on the geometry of the real configuration space will
be very relevant for the Minkowskian quantum theory.

Another key set of issues that has not been explored here are
global questions.  As quantum states will be functionals on the
whole configuration space, it is essential that more be learned
about the global topological and differential properties of
${\cal C}_{diff}$.  We may note that the existence of a preferred
function such as $\int Y_{CS}$ on this space may play an important
role in this analysis.

Given those properties of the Chern-Simons function as a coordinate
on the configuration space that we were able to discover, we went
on to study the question of whether the Chern-Simons functional
might play a useful role as a time parameter in the quantum
theory.
The results we found here are necessarily all, in one way or another
incomplete.  However, we found two ways in which
the Chern-Simons function might play a significant  role in the quantum
theory.

iv)  It emerges in the semiclassical limit of the Hamiltonian
constraint as the natural time coordinate
for quantum gravity coupled to matter, if the
cosmological constant is nonvanishing and the vacuum is
taken to be the Chern-Simons state (Section 5).  It is
interesting to note also that in this
limit, the Chern-Simons form reduces to its
imaginary part, which is in turn proportional to the  trace
of the extrinsic curvature.   This suggest that in the
physical, Minkowskian case, it is the imaginary part of the
Chern-Simons invariant that is playing the role of time.  However,
unlike the case of the Euclideanized theory, we presently have
good evidence for this only at the semiclassical level.

v)  $\tau_{CS}$
provides a periodic coordinate on the Euclidean section
of the configuration space.  This means that to the extent that
a quantum theory of gravity can be established as a theory of
motion on this configuration space, the theory must
be intrinsically thermal, in that physical expectation values
of the Euclideanized theory
must satisfy the $KMS$
condition when expressed in terms of the Chern-Simons time.

In the end, even if
it turns out to be possible to invent a form
of quantum cosmology that makes sense when evolution is
described with respect to any particular clock, it will certainly
be the case that important features of the theory will be more
transparent when studied with one choice of time than
another.  We believe that the results we have reported here
suggest that whatever the outcome of the deep problem of time
in quantum cosmology, the Chern-Simons invariant can play
a significant role in our unraveling the dynamics of general
relativity, at both the classical and quantum level.

\section*{Acknowledgements}

We would like to thank Abhay Ashtekar, Lay Nam Chang,
Louis Crane, Chris
Isham,
Ted Jacobson,  Karel Kuchar,
Guillermo A. Mena Marugan, Vince Moncrief,
Jorge Pullin,
Carlo Rovelli and Richard Woodard for helpful
discussions. This work has been supported by the National Science
Foundation under grant NSF-PHY93-96246 and research funds
provided by the Pennsylvania State University.

\end{document}